\documentclass[journal]{IEEEtran}

\bstctlcite{IEEEexample:BSTcontrol}

\usepackage{epsfig,rotating,setspace,latexsym,amsmath,epsf,amssymb,amsfonts,bm,theorem,cite,authblk, bbm,color, hyperref, multirow, caption, graphicx, subcaption, url, xcolor}

\IEEEoverridecommandlockouts
\allowdisplaybreaks

\title{Semantic Leakage and Privacy Preservation in Relay-Assisted Semantic Communications}

\author[1]{Yalin E. Sagduyu}
\author[1]{Tugba Erpek}
\author[2]{Aylin Yener}
\author[3]{Sennur Ulukus}

\affil[1]{\normalsize  Nexcepta, Gaithersburg, MD, USA}
\affil[2]{\normalsize  The Ohio State University, Columbus, OH, USA}
\affil[3]{\normalsize  University of Maryland, College Park, MD, USA}

\begin{document}
\bstctlcite{BSTcontrol}

\pagestyle{empty}
\maketitle
\thispagestyle{empty}

\begin{abstract}
Semantic communication (SemCom) has emerged as a promising paradigm in which the transmission of task-relevant information is prioritized over raw data, enabling efficient and robust communication under resource and channel constraints. In this paper, the privacy implications of relay-assisted SemCom systems are studied, where the intermediate relay node operates directly on learned latent representations. It is shown that the relay, even without access to source data, can reliably infer semantic meaning and reconstruct signals with performance comparable to that of the legitimate receiver, revealing a fundamental privacy vulnerability of semantic representations. To address this issue, an iterative adversarial training framework is proposed in which a strong, adaptively trained eavesdropper at the relay is explicitly accounted for. The proposed approach alternates between optimizing the relay's eavesdropping function and the legitimate system, resulting in representations that preserve semantic decoding performance at the intended receiver while degrading semantic inference at the relay. The semantic accuracy gap between the legitimate receiver and the eavesdropper is significantly enlarged across channel conditions. Importantly, this protection is achieved in a stealthy manner, with high reconstruction fidelity maintained while semantic leakage is selectively suppressed.
\end{abstract}


\section{Introduction}

Conventional communication systems are designed to reliably transmit bits, whereas \emph{semantic communication} (SemCom) shifts the focus toward conveying task-relevant meaning \cite{guler2014, gunduz2022beyond}. By leveraging deep neural networks (DNNs) to encode and decode semantic representations, such systems can significantly improve robustness, compression efficiency, and adaptability to varying channel conditions \cite{huang2022toward, xie2021deep, sagduyu2024will}. 

\emph{Relay-assisted communication} is essential to extend coverage, improve reliability, and overcome limitations of direct links between the source-destination pairs. Integrating relays into SemCom systems is therefore a natural extension, enabling cooperative transmission of learned representations \cite{lin2024semantic, arda2024semantic, guo2024distributed, luo2022autoencoder}. When applied to SemCom, relays can operate directly on latent representations rather than decoded signals, allowing for flexible and efficient forwarding strategies.

However, this architectural shift introduces a critical and largely unexplored \emph{privacy challenge}. Unlike conventional relays that forward encoded bits without semantic interpretation, semantic relays inherently process representations that embed high-level information. As a result, an \emph{untrusted relay} can act as an intelligent eavesdropper, capable of learning and inferring semantic meaning directly from the observed signals. Controlling this semantic leakage at intermediate nodes is a key challenge for SemCom.

Security and privacy in SemCom have emerged as critical research directions, as the shift from bit-level transmission to representation-level encoding fundamentally changes the attack surface \cite{sagduyu2023semantic, sagduyu2023task, yang2024secure}. Unlike conventional systems, where confidentiality is enforced through bit-level encryption, SemCom embeds high-level information directly into transmitted representations, making them inherently vulnerable to inference attacks. Privacy-preserving methods based on adversarial perturbations, representation obfuscation, and multi-task learning have been investigated \cite{liu2023semprotector, luo2023encrypted, wang2024privacy}. However, these methods focus on point-to-point settings and do not consider intermediate nodes. In relay-assisted systems, the relay observes a highly informative latent representation and can act as a powerful, learnable eavesdropper, exposing major privacy threats.

\begin{figure}[t!]
    \centering
    \includegraphics[width=\linewidth]{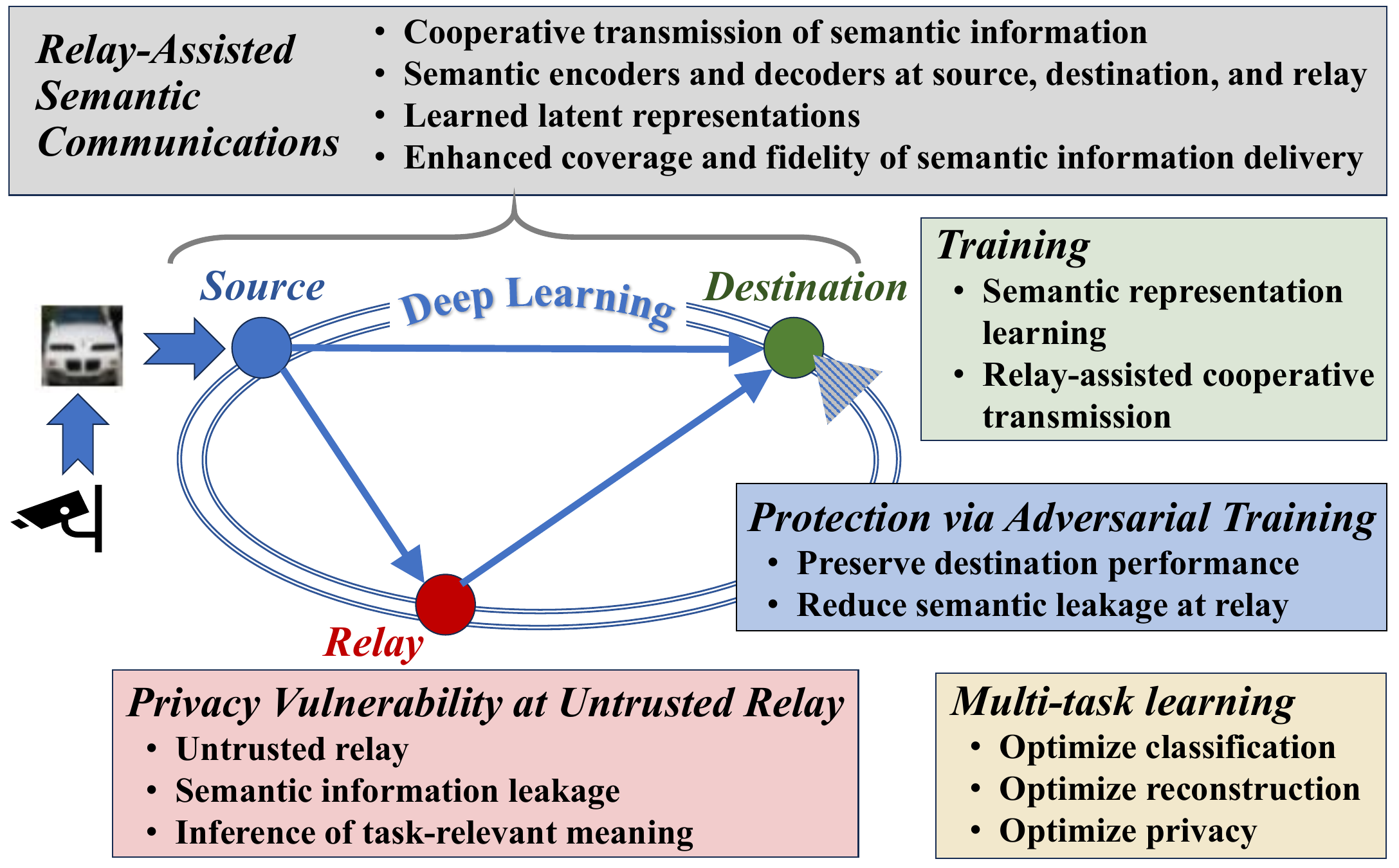}
    \caption{Relay-assisted SemCom with privacy vulnerability and protection.}
    \label{fig:semrelayprivacy}
\end{figure}

Fig.~\ref{fig:semrelayprivacy} illustrates relay-assisted SemCom, where cooperative relaying improves semantic delivery while an untrusted relay introduces semantic privacy vulnerabilities. In this paper, we study privacy preservation in relay-assisted SemCom, where the relay is modeled as a strong, learnable eavesdropper operating on latent representations. We first show that the relay effectively infers semantic meaning from its observations, revealing a significant privacy risk. Specifically, the relay can be trained to achieve semantic classification and reconstruction performance comparable to that of the legitimate receiver, even under compressed representations. This reveals a key SemCom vulnerability where robustness to channel impairments facilitates \emph{semantic leakage}.

Conventional physical-layer security methods such as transmit power control and source-side artificial noise injection have fundamental limitations in relay-assisted SemCom because the relay observes a stronger first-hop semantic representation than the destination and can therefore still reliably infer semantic information. To address the privacy challenge introduced by the untrusted relay, we propose an iterative \emph{adversarial training} framework that explicitly models the relay as an adaptive eavesdropper. The idea is to alternate between training the eavesdropper to maximize its inference capability and training the legitimate system to degrade the eavesdropper's performance while preserving its own utility. This leads to a selectively informative representation that preserves semantic accuracy at the intended receiver while degrading semantic separability at the relay. The resulting protection is inherently stealthy, suppressing semantic leakage without introducing noticeable degradation in reconstruction fidelity. 

The remainder of the paper is organized as follows. Sec.~\ref{sec:semcomprivacy} presents the relay-assisted SemCom design and its privacy vulnerability. Sec.~\ref{sec:iterative} introduces the iterative adversarial framework for privacy protection. Sec.~\ref{sec:conclusion} concludes the paper.

\section{Semantic Communications with an Eavesdropping Relay  \label{sec:semcomprivacy}}

\subsection{Relay-Assisted Semantic Communication System Model}

\begin{figure}[t!]
  \centering
    \includegraphics[width=\linewidth]{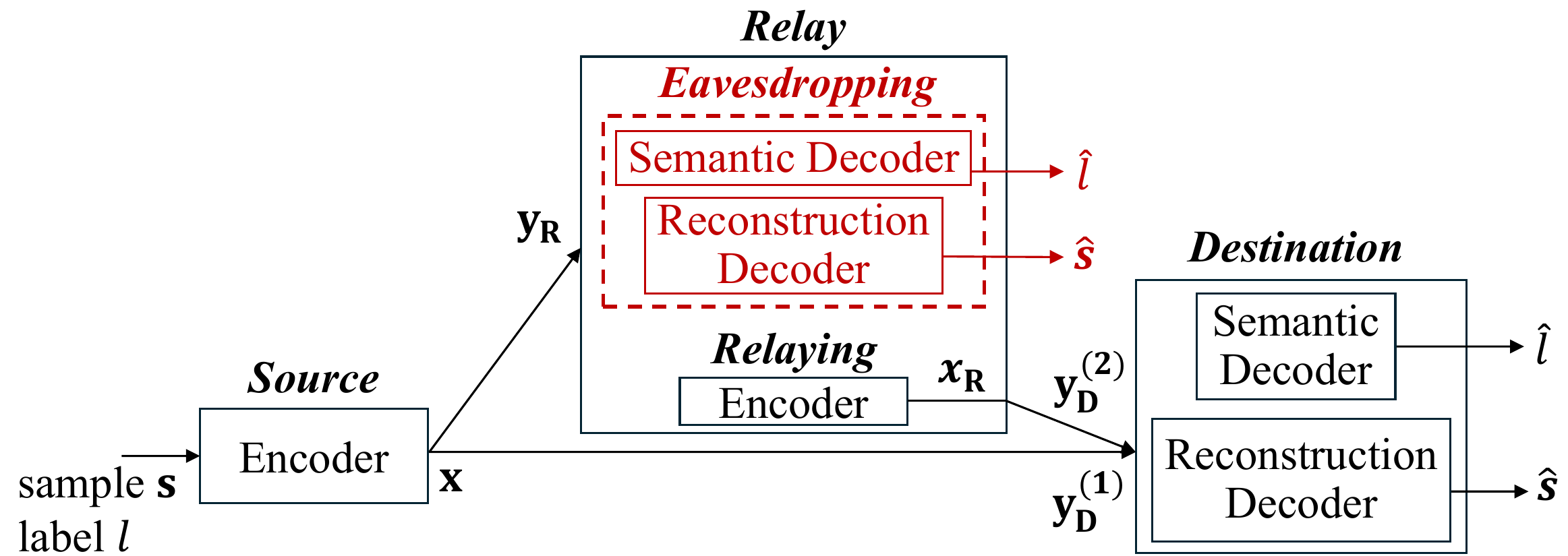}
    \caption{System model of semantic relay communications.}\label{fig:semrelaycom}
\end{figure}

We consider a relay-assisted SemCom system with a source communicating with a destination through the assistance of a relay, as shown in Fig.~\ref{fig:semrelaycom}. Let $\mathbf{s}$ denote the input data sample, i.e., an image with pixel values normalized to $[0,1]$, with $C$ channels, height $H$, and width $W$. For the CIFAR-10 dataset considered here, $C \!=\! 3$, $H \!=\! 32$, and $W \!=\!32$. Let $l$ denote the associated semantic label. There are $L = 10$ classes for CIFAR-10. The transmitter maps the input sample $\mathbf{s}$ into a complex-valued transmitted latent signal $\mathbf{x}$ of length $N$, where $N$ denotes the number of complex channel uses and controls the semantic bandwidth/compression ratio. In implementation, this signal is represented using a real-valued tensor with two channels and length $N$, corresponding to its in-phase and quadrature components, which together form the equivalent complex-valued representation.

The input sample $\mathbf{s}$ is processed by the encoder DNN (parameterized by $\theta_A$) as $f_{\theta_A}(\mathbf{s})$. The output signal $\mathbf{x}$ is normalized per sample to satisfy a unit average transmit power constraint. The encoder $f_{\theta_A}$ is a convolutional neural network that extracts hierarchical spatial features and compresses them into a latent representation using two convolutional stages. The first stage applies two $3\times3$ convolutions mapping $C \rightarrow 64 \rightarrow 64$, each followed by batch normalization and ReLU, then a $2\times2$ max-pooling layer (stride 2) and dropout with probability $0.25$. The second stage applies two $3\times3$ convolutions mapping $64 \rightarrow 128 \rightarrow 128$, again followed by batch normalization and ReLU, then a $2\times2$ max-pooling layer (stride 2) and dropout with probability $0.25$. The resulting feature tensor has size $128 \times (H/4) \times (W/4)$ (for CIFAR-10: $128 \times 8 \times 8$), which is flattened and passed through fully connected layers $128(H/4)(W/4) \rightarrow 1024 \rightarrow 2N$ with ReLU activation. The output is reshaped into a two-channel signal representing the real and imaginary parts of $\mathbf{x}$; for $N=256$, this corresponds to 512 real-valued components.

The wireless channel is modeled as block Rayleigh fading with additive white Gaussian noise (AWGN). For any transmitted signal $\mathbf{x}$, the received signal is given by $\mathbf{y} = h \mathbf{x} + \mathbf{n}$, where the channel coefficient $h$ is a complex scalar that remains constant across the $N$ symbols of a given sample, $\mathbf{n}$ denotes the noise, and $\mathbf{y}$, $\mathbf{x}$, and $\mathbf{n}$ are complex-valued signals of length $N$. 

In relay-assisted operation, communication proceeds in two phases.
\begin{enumerate}
\item
In the first phase, the source transmits $\mathbf{x}$, and the destination and relay receive signals $\mathbf{y}_D^{(1)}$ and $\mathbf{y}_R^{(1)}$, respectively.

\item
In the second phase, the relay processes $\mathbf{y}_R^{(1)}$ through the mapping $g_{\theta_R}(\cdot)$ and transmits $\mathbf{x}_R$, and the destination receives the second-phase signal $\mathbf{y}_D^{(2)}$. The destination then forms the joint observation $\mathbf{y}_D$ by concatenating $\mathbf{y}_D^{(1)}$ and $\mathbf{y}_D^{(2)}$, resulting in a representation whose dimension is twice that of each individual phase observation.
\end{enumerate}

The relay mapping $g_{\theta_R}$ maps a latent signal to another one of the same length without reconstructing the input sequence. The relay is implemented as a one-dimensional convolutional residual network that preserves the sequential structure of the signal-level latent representation. The input is treated as a two-channel sequence of length $N$, and an initial 1D convolution (kernel size 3, padding 1) maps 2 channels to 128 channels. Two residual blocks follow, each consisting of two 1D convolutional layers (kernel size 3, padding 1, 128 channels) with ReLU activations and identity skip connections. A final 1D convolution (kernel size 3, padding 1) maps the representation back to 2 channels. The relay output is normalized per sample to satisfy the transmit power constraint.

At the receiver, the destination applies two decoders to the joint observation $\mathbf{y}_D$. The reconstruction decoder $d_{\theta_D}^{\mathrm{rec}}$ maps the received signal to a reconstructed image $\hat{\mathbf{s}}$. Internally, the input is represented in real form and flattened to dimension $4N$, then passed through fully connected layers $4N \rightarrow 1024 \rightarrow 256(H/4)(W/4)$. The output is reshaped into a tensor of size $256 \times (H/4) \times (W/4)$, followed by two transposed convolution layers with kernel size 4, stride 2, and padding 1, mapping $256 \rightarrow 128 \rightarrow 64$ channels, each followed by batch normalization and ReLU. A final $3\times 3$ convolution maps 64 channels to $C$, and a sigmoid activation produces $\hat{\mathbf{s}}$. The classification decoder $d_{\theta_D}^{\mathrm{cls}}$ maps the same input to semantic prediction logits $\mathbf{z}_D$ over $L$ classes. It consists of fully connected layers $4N \rightarrow 512 \rightarrow 256 \rightarrow L$, with ReLU activations and dropout with probability $0.5$ after the first two layers. The predicted label $\hat{l}_D$ corresponds to the class associated with the maximum component of $\mathbf{z}_D$.

\subsection{Training of Legitimate SemCom Relay System}

SemCom is trained in two stages. In Stage 1, the parameters $(\theta_A, \theta_R, \theta_D^{\mathrm{rec}}, \theta_D^{\mathrm{cls}})$ are jointly optimized using a weighted objective that combines reconstruction fidelity and semantic classification performance. The overall loss consists of four terms weighted by $\lambda_{\mathrm{rec}}$, $\lambda_{\mathrm{ssim}}$, $\lambda_{\mathrm{psnr}}$, and $\lambda_{\mathrm{ce}}$.

\begin{enumerate}
\item The first term is the mean squared error (MSE), denoted as $\mathrm{MSE}(\hat{\mathbf{s}}, \mathbf{s})$, which measures the average squared difference between the reconstructed sample $\hat{\mathbf{s}}$ and the original sample $\mathbf{s}$ over all image elements. Lower MSE indicates improved reconstruction fidelity.

\item The second term is the structural similarity index, denoted as $\mathrm{SSIM}(\hat{\mathbf{s}}, \mathbf{s})$, which evaluates perceptual similarity between $\hat{\mathbf{s}}$ and $\mathbf{s}$ by comparing their luminance, contrast, and structural consistency. Since higher SSIM values correspond to better perceptual similarity, the optimization minimizes $1-\mathrm{SSIM}(\hat{\mathbf{s}}, \mathbf{s})$.

\item The third term is the peak signal-to-noise ratio, denoted as $\mathrm{PSNR}(\hat{\mathbf{s}}, \mathbf{s})$, which measures reconstruction quality in decibels based on the reconstruction error. Larger PSNR indicates improved reconstruction quality. The optimization objective therefore encourages higher PSNR values during training.

\item The fourth term is the categorical cross-entropy loss, denoted as $\mathcal{L}_{\mathrm{CE}}(\mathbf{z}_D,l)$, where $\mathbf{z}_D$ represents the semantic prediction logits at the destination and $l$ denotes the ground-truth semantic label. This term measures the semantic classification error and encourages accurate semantic inference at the receiver.
\end{enumerate}
For performance evaluation, weights for loss terms are selected as $\lambda_{\mathrm{rec}} = 5$, $\lambda_{\mathrm{ssim}} = 2$, $\lambda_{\mathrm{psnr}} = 1$, and $\lambda_{\mathrm{ce}} = 1$.
The MSE term enforces pixel-wise fidelity and stabilizes optimization, while the PSNR term introduces a logarithmic sensitivity to reconstruction error that emphasizes improvements in high-fidelity regimes. The SSIM term further promotes structural consistency in the reconstructed images.

\subsection{Eavesdropping Relay Model}
The relay simultaneously acts as a forwarding node for cooperative communication and as an untrusted eavesdropper that attempts semantic inference from its received observation. In Stage 2 of SemCom training, the legitimate system is frozen. The first-phase relay observation, denoted by $\mathbf{y}_R^{(1)}$, is a complex-valued signal of length $N$. This observation serves as the input $\mathbf{y}_E$ to the relay’s eavesdropping function and is processed through the relay’s reconstruction and classification decoders for semantic eavesdropping, as shown in Fig.~\ref{fig:semrelaycom}. The reconstruction decoder $d_{\theta_E}^{\mathrm{rec}}$ follows the same architectural structure as the destination's reconstruction decoder but operates on a latent signal of length $N$. The classification decoder $d_{\theta_E}^{\mathrm{cls}}$ produces semantic prediction logits $\mathbf{z}_E$ over the $L$ classes using fully connected layers of dimensions $2N \rightarrow 512 \rightarrow 256 \rightarrow L$, with ReLU activations and dropout with probability $0.5$ applied after the first two layers.

The eavesdropping relay is trained using a joint loss function that considers both semantic inference performance and signal reconstruction quality. The first component is a reconstruction loss weighted by $\lambda_{E,\mathrm{rec}}$, which measures the squared difference between the reconstructed sample $\hat{\mathbf{s}}_E$ and the original sample $\mathbf{s}$. Here, $\hat{\mathbf{s}}_E$, generated as $d_{\theta_E}^{\mathrm{rec}}(\mathbf{y}_E)$, denotes the reconstruction produced by the eavesdropper reconstruction decoder from the received observation $\mathbf{y}_E$. Minimizing this term encourages the eavesdropper to accurately reconstruct the transmitted sample. The second component is a semantic classification loss weighted by $\lambda_{E,\mathrm{cls}}$, given by the categorical cross-entropy loss $\mathcal{L}_{\mathrm{CE}}(\mathbf{z}_E,l)$, where $\mathbf{z}_E$ represents the semantic prediction logits generated by the eavesdropper and $l$ denotes the ground-truth semantic label. Minimizing this term improves the semantic classification capability of the eavesdropper. For performance evaluation, the reconstruction and semantic classification losses are weighted by $\lambda_{E,\mathrm{rec}}\!=\!1$ and $\lambda_{E,\mathrm{cls}}\!=\!1$, respectively, giving equal importance to reconstruction fidelity and semantic inference performance during eavesdropper training.
Only the parameters $(\theta_E^{\mathrm{rec}}, \theta_E^{\mathrm{cls}})$ are optimized in this stage.

\subsection{Relay-assisted SemCom Performance and Eavesdropping Vulnerability}

For performance evaluation, training and evaluation are performed on NVIDIA RTX PRO 6000 Blackwell GPUs using batch size 128, learning rate $10^{-3}$, and 1000 training epochs for each stage. The source encoder, destination decoder system, forwarding relay, and relay-eavesdropper model contain 9.2M, 19.8M, 0.2M, and 18.8M parameters for $N=256$, and 8.7M, 18.0M, 0.2M, and 17.9M parameters for $N=32$. The channel is modeled as Rayleigh fading with AWGN under varying SNR conditions. In this framework, the encoder learns a semantically informative latent representation $\mathbf{x}$. Since no privacy constraint is imposed during Stage 1, the relay observation $\mathbf{y}_E$ retains sufficient semantic information, enabling the relay to learn both reconstruction and classification mappings that can achieve performance comparable to that of the legitimate receiver.

\begin{figure*}[t]
    \centering
    \begin{subfigure}{0.32\textwidth}
        \centering
        \includegraphics[width=\linewidth]{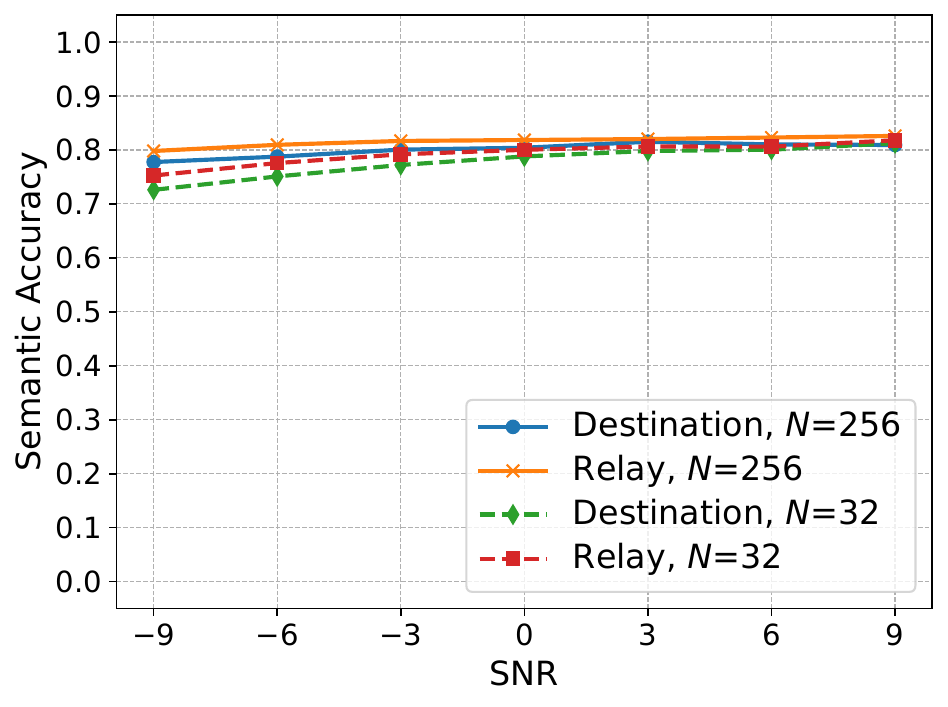}
        \caption{Semantic accuracy.}
    \end{subfigure}
    \hfill
    \begin{subfigure}{0.32\textwidth}
        \centering
        \includegraphics[width=\linewidth]{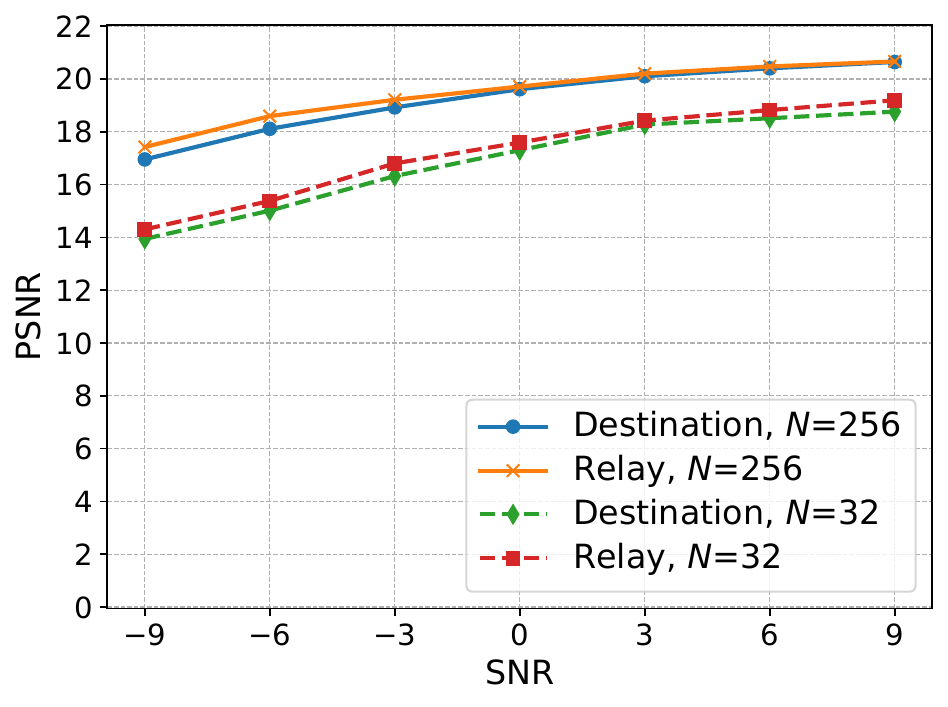}
        \caption{PSNR.}
    \end{subfigure}
    \hfill
    \begin{subfigure}{0.32\textwidth}
        \centering
        \includegraphics[width=\linewidth]{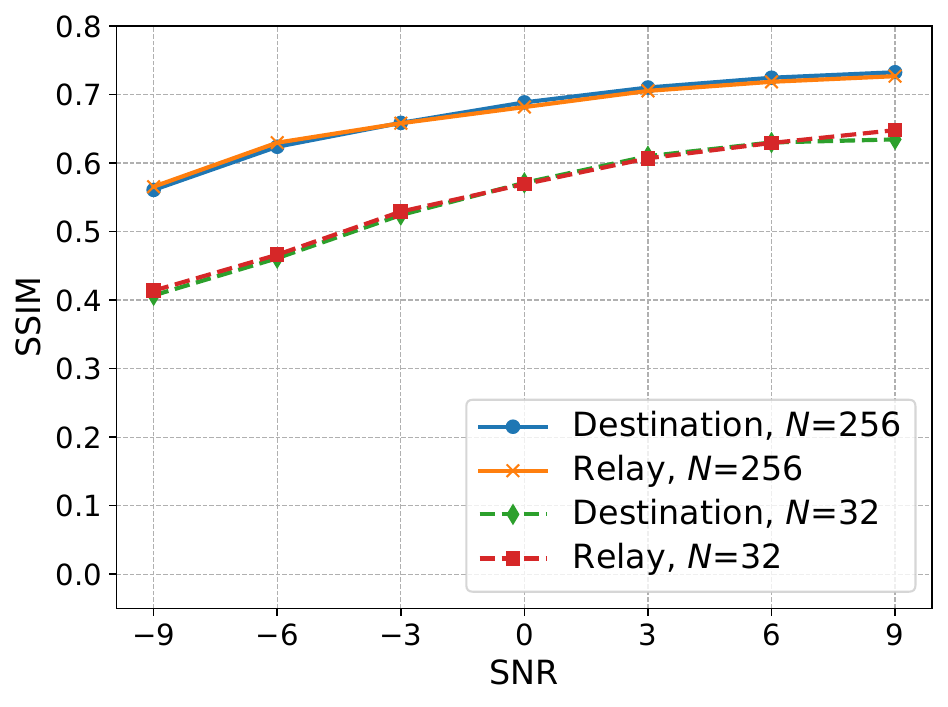}
        \caption{SSIM.}
    \end{subfigure}
    \caption{Performance (semantic accuracy, PSNR, and SSIM) of the destination and eavesdropping relay across SNR for different latent dimensions $N$.}
    \label{fig:baseline_all_metrics}
\end{figure*}

We evaluate the SemCom system across a range of channel conditions, with performance illustrated in terms of semantic classification accuracy, PSNR, and SSIM for both the legitimate receiver (destination) and the eavesdropper (relay). The corresponding trends are shown in Fig.~\ref{fig:baseline_all_metrics}, where each subfigure depicts one metric as a function of the signal-to-noise ratio (SNR) for two latent dimensions, $N=256$ and $N=32$.

The semantic classification accuracy in Fig.~\ref{fig:baseline_all_metrics}(a) increases with SNR, reflecting improved performance under favorable channel conditions. This trend holds for both latent dimensions. Increasing the latent dimension from $N=32$ to $256$ improves performance for both the destination and relay, as higher-dimensional representations preserve more semantic information. Even at $N\!=32$, the system maintains reasonable accuracy across SNR values, indicating effective semantic transmission under strong compression. In both cases, the relay's accuracy closely tracks and occasionally exceeds that of the destination because it does not experience the second-hop impairment, directly observes the first-hop representation, and operates on representations that are not optimized to distinguish legitimate from unintended receivers.

The reconstruction performance, measured by PSNR in Fig.~\ref{fig:baseline_all_metrics}(b), follows a similar trend, increasing monotonically with SNR for both receivers and both values of $N$. A larger latent dimension yields consistently better reconstruction quality, reflecting increased representational capacity. However, the gap between $N=32$ and $N=256$ remains moderate, indicating that even compact representations preserve a substantial portion of the signal structure. Notably, the PSNR performance for the destination and relay remains closely aligned across both values of $N$, showing that the relay retains access to high-fidelity reconstruction even at lower dimensions.

The SSIM results in Fig.~\ref{fig:baseline_all_metrics}(c) show a similar trend. SSIM increases with SNR and latent dimensions, reflecting improved preservation of perceptual structures such as edges and textures. While $N=256$ yields higher SSIM overall, performance at $N=32$ remains strong, indicating that perceptual quality is preserved under aggressive dimensionality reduction. The relay's SSIM closely tracks that of the destination across all conditions, confirming that perceptual information is largely preserved in the relay observation.

Taken together, Fig.~\ref{fig:baseline_all_metrics} shows that increasing the latent dimension improves both semantic and reconstruction performance, while even reduced dimensions maintain reasonable performance. The close alignment between the destination and relay across all metrics indicates that semantic and structural information remain readily accessible at the relay. This exposes a fundamental vulnerability of SemCom, namely that reducing the representation size alone does not prevent semantic leakage. Even compressed representations retain sufficient information for unintended receivers to recover semantic meaning and signal structure, motivating the need for privacy-aware mechanisms such as adversarial training.

Conventional physical-layer security schemes such as transmit power control and source-side artificial noise injection do not provide effective privacy protection in relay-assisted SemCom. Since the relay observes a stronger first-hop signal than the destination, these methods degrade both receivers without selectively suppressing relay inference and therefore do not help increase the destination-relay semantic accuracy gap, as also confirmed by our performance evaluation. This highlights the need for privacy-aware representation learning in relay-assisted SemCom.

\section{Iterative Adversarial Training for Protection against Untrusted Relay \label{sec:iterative}}

\subsection{Alternating Adversarial Optimization}

To mitigate semantic leakage, an iterative adversarial training framework is introduced in which the legitimate Source-Relay-Destination (SRD) system and the eavesdropper (relay) are updated in an alternating manner. The procedure follows a sequential min-max update rule, where the relay is first trained against a fixed SRD system, and then SRD is updated against a fixed relay. This alternating process is repeated throughout training. Let $\Theta$ denote the parameters of the legitimate SRD system, including the source, forwarding relay, and destination, and let $\Phi$, consisting of $\theta_E^{\mathrm{rec}}$ and $\theta_E^{\mathrm{cls}}$, denote the parameters of the eavesdropping function at the relay (as introduced in Stage 2), corresponding to its reconstruction and classification networks. 

At iteration $t$, the training consists of two steps.

\begin{enumerate}
\item In the first step of iterative optimization, the SRD parameters, denoted by $\Theta$, are held fixed while the eavesdropping relay parameters, denoted by $\Phi$, are updated to minimize the relay's own inference loss. At iteration $t$, the relay parameters are updated from $\Phi^{(t)}$ to $\Phi^{(t+1)}$ by minimizing the eavesdropping objective $\mathcal{L}_E(\Theta^{(t)},\Phi)$. The eavesdropping loss consists of two components. 
\begin{enumerate}
    \item The first component is a reconstruction loss weighted by $\lambda_{E,\mathrm{rec}}$, which measures the fidelity of the reconstructed sample $\hat{\mathbf{s}}_E$ relative to the original sample $\mathbf{s}$. Minimizing this term encourages the relay to improve its reconstruction capability, enabling evaluation of stealthy protection.
    \item The second component is the categorical cross-entropy loss weighted by $\lambda_{E,\mathrm{cls}}$, denoted as $\mathcal{L}_{\mathrm{CE}}(\mathbf{z}_E,l)$, where $\mathbf{z}_E$ represents the semantic prediction logits generated by the eavesdropping relay and $l$ denotes the ground-truth semantic label. Minimizing this term improves the semantic classification capability of the eavesdropping relay.
\end{enumerate}

Together, these two terms define the eavesdropping loss $\mathcal{L}_E(\Theta,\Phi)$ and enable the relay to optimize both semantic inference and reconstruction fidelity while the SRD model remains fixed. In practice, the semantic classification term dominates, reflecting the goal of maximizing the relay's inference capability. This step ensures that the relay acts as a strong adaptive adversary for the current latent representation.

\item In the second step of iterative optimization, the eavesdropping function of the relay is held fixed while the SRD system parameters, denoted by $\Theta$, are updated using an adversarially augmented training objective. At iteration $t$, the SRD parameters are updated from $\Theta^{(t)}$ to $\Theta^{(t+1)}$ by minimizing an adversarially augmented objective consisting of the legitimate loss $\mathcal{L}_{\mathrm{SRD}}(\Theta)$ and an adversarial penalty term based on the relay's semantic classification loss.
\begin{enumerate}
    \item The first component, denoted by $\mathcal{L}_{\mathrm{SRD}}(\Theta)$, represents the standard training loss of the legitimate SRD communication system and combines reconstruction fidelity and semantic classification objectives for the intended receiver.
    \item The second component is proportional to the eavesdropping relay's semantic classification loss, denoted as $\mathcal{L}_{\mathrm{CE}}(\mathbf{z}_E,l)$, where $\mathbf{z}_E$ represents the semantic prediction logits generated by the relay and $l$ denotes the ground-truth semantic label. This term is weighted by the positive coefficient $\gamma_{\mathrm{eve}}$, which controls the strength of the adversarial penalty.
\end{enumerate}

Because the relay's semantic classification loss is subtracted from the legitimate objective, the SRD system is encouraged to increase the relay's classification error and reduce its ability to infer semantic labels from the transmitted signal. At the same time, it continues optimizing the legitimate communication objective to preserve reliable semantic inference and reconstruction at the destination, thereby improving privacy while maintaining communication performance.

\end{enumerate}

This alternating update induces a competitive dynamic between SRD and the relay. The relay continuously improves its ability to extract semantic information from the relay observation by minimizing $\mathcal{L}_E$, while SRD adapts by reshaping the transmitted representation to increase the relay's classification error while maintaining its own performance. 

\subsection{Gradient-Based Training Dynamics}

At the gradient-update level, the optimization proceeds by alternately updating the eavesdropping relay parameters $\Phi$ and the SRD system parameters $\Theta$ using gradient descent. The relay parameters $\Phi$ are updated using the gradient of the eavesdropping loss $\mathcal{L}_E(\Theta,\Phi)$ with respect to $\Phi$. The update step is scaled by the learning rate $\eta_E$, which controls the adaptation rate of the eavesdropping relay during training. This update encourages the relay to improve both semantic inference accuracy and reconstruction quality.

The SRD system parameters $\Theta$ are updated using the gradient of an adversarially augmented objective consisting of the legitimate communication loss $\mathcal{L}_{\mathrm{SRD}}(\Theta)$ and the adversarial semantic classification term weighted by $\gamma_{\mathrm{eve}}$. The update step is scaled by the learning rate $\eta_M$, which controls the adaptation rate of the SRD system. Because the eavesdropping semantic classification loss $\mathcal{L}_{\mathrm{CE}}(\mathbf{z}_E,l)$ is subtracted from the legitimate objective, the gradient update encourages the SRD system to increase the relay's semantic classification error while still maintaining reliable semantic inference and reconstruction performance for the intended receiver. This alternating optimization reduces the relay's semantic inference capability. Importantly, the adversarial term in the SRD update acts only on the relay's semantic classification loss, rather than its full reconstruction objective. As a result, the training process specifically targets the degradation of semantic inference at the relay, without explicitly suppressing reconstruction quality. This design aligns with the primary goal of enlarging the semantic accuracy gap between the destination and the relay.

Overall, the iterative framework performs alternating adversarial optimization in which the relay adapts to the current transmission strategy while SRD counter-adapts to reduce semantic leakage. Through repeated updates, the learned representation remains discriminative for the legitimate receiver while becoming less separable for the eavesdropper, thereby widening the destination-relay semantic accuracy gap.

\subsection{Learning Dynamics and Semantic Accuracy Gap}
Next, we evaluate the effectiveness of the proposed iterative adversarial training framework in improving resistance to eavesdropping. The results are presented for multiple channel conditions, with performance illustrated as a function of the weight assigned to the eavesdropping protection loss, denoted by $\gamma_{\mathrm{eve}}$.We adopt $N=256$ for the results in this section. Both the SRD system and the eavesdropping relay are trained using the Adam optimizer with learning rates $\eta_M=\eta_E=10^{-3}$. We report the semantic classification accuracy at the destination, the semantic accuracy at the eavesdropping relay, and the resulting destination-relay accuracy gap.

\begin{figure}[t!]
    \centering
    \includegraphics[width=0.8\linewidth]{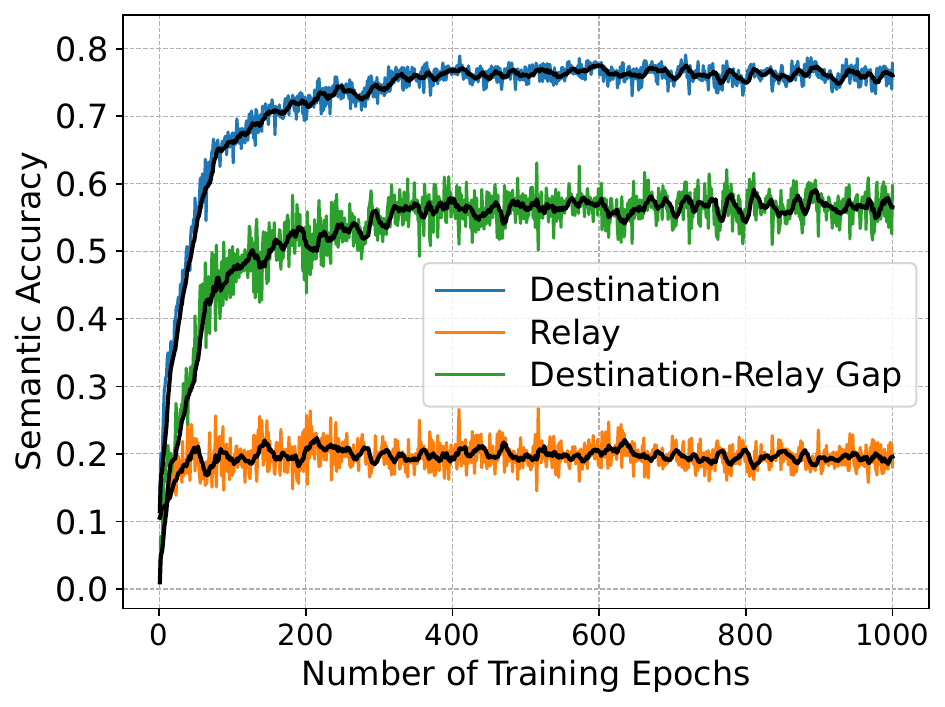}
    \caption{Evolution of semantic accuracy at the destination and the relay during iterative adversarial training when SNR = $6$ dB and  $\gamma_{\mathrm{eve}} = 5$.}
    \label{fig:training_evolution}
\end{figure}

We first examine the temporal evolution of the adversarial training process. Fig.~\ref{fig:training_evolution} shows the semantic accuracy at the destination and the relay over training epochs for a representative setting with SNR $= 6$ dB and a moderate adversarial weight $\gamma_{\mathrm{eve}} = 5$. At the early stages of training, the destination and the relay exhibit similar performance, reflecting the absence of adversarial separation in the initial learned representation. As training progresses, a clear divergence emerges: the relay's accuracy gradually decreases while the destination's accuracy stabilizes at a higher level. This behavior reflects the alternating training mechanism in which the relay improves inference capability while the legitimate system suppresses relay predictions. Over successive iterations, this interaction widens the semantic accuracy gap between the destination and the relay. The convergence pattern in Fig.~\ref{fig:training_evolution} indicates a stable equilibrium where the destination maintains high accuracy while the relay's capability is suppressed, demonstrating that the adversarial objective progressively reshapes the latent representation to reduce semantic leakage.

\begin{figure*}[t]
    \centering
    \begin{subfigure}{0.32\textwidth}
        \centering
        \includegraphics[width=\linewidth]{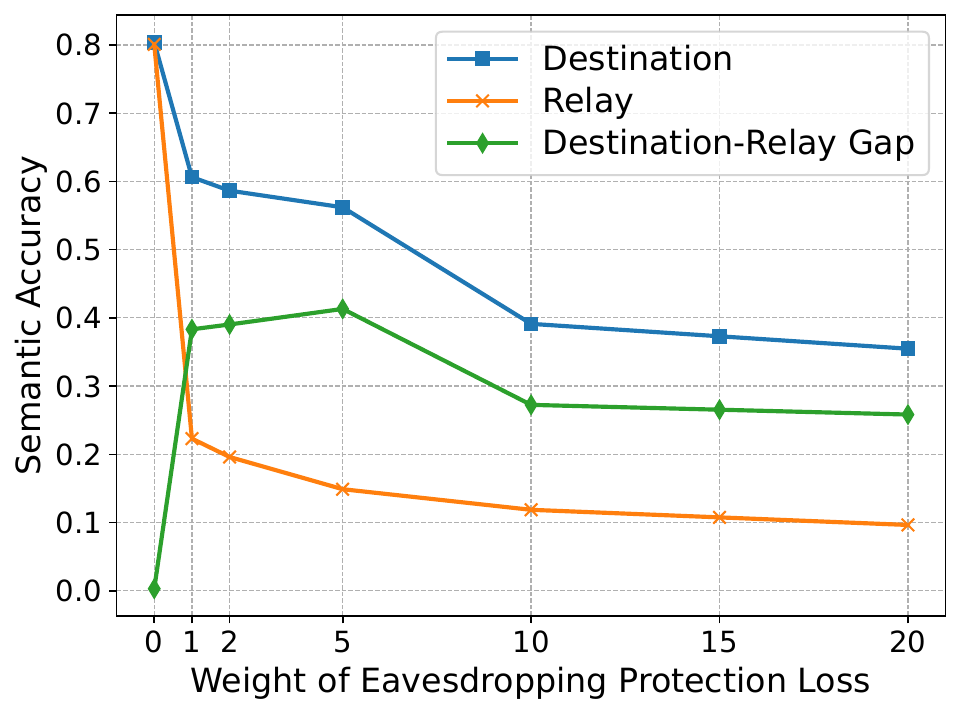}
        \caption{SNR = $-6$ dB.}
    \end{subfigure}
    \hfill
    \begin{subfigure}{0.32\textwidth}
        \centering
        \includegraphics[width=\linewidth]{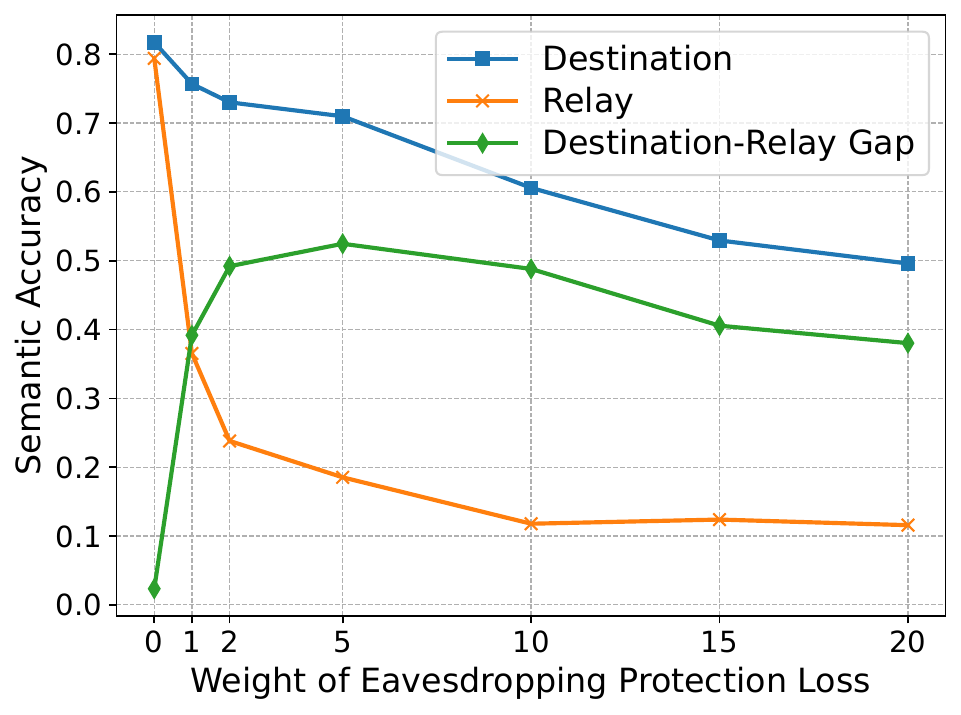}
        \caption{SNR = $0$ dB.}
    \end{subfigure}
    \hfill
    \begin{subfigure}{0.32\textwidth}
        \centering
        \includegraphics[width=\linewidth]{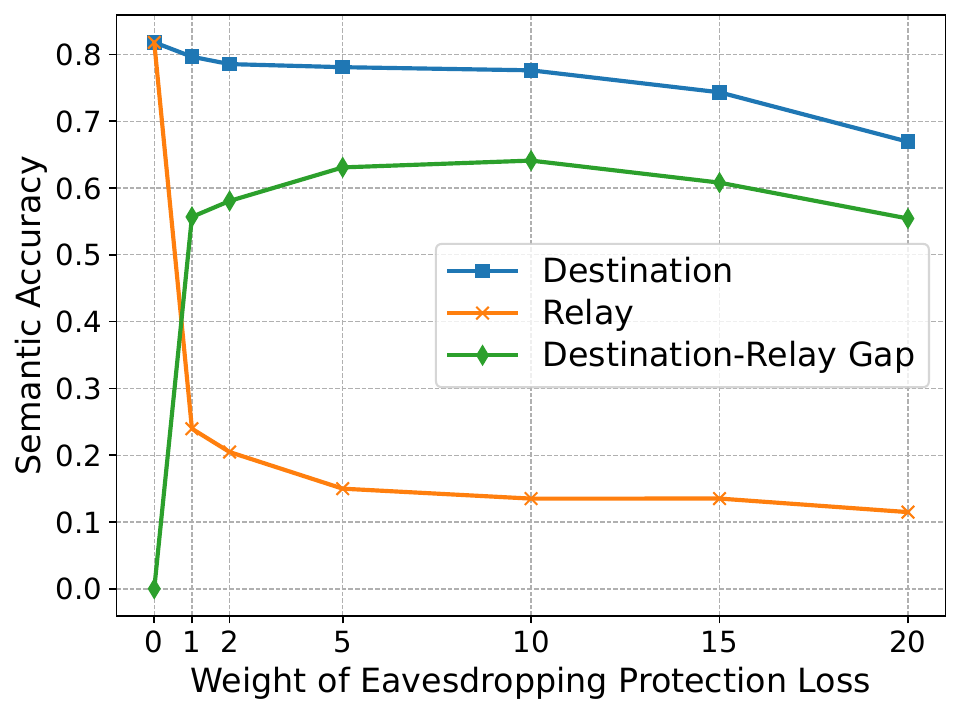}
        \caption{SNR = $6$ dB.}
    \end{subfigure}
    \caption{Effect of adversarial loss weight $\gamma_{\mathrm{eve}}$ on semantic accuracy at the destination and the relay, and the resulting accuracy gap.}
    \label{fig:semantic_gap_all}
\end{figure*}

We next examine the effect of the adversarial loss weight $\gamma_{\mathrm{eve}}$ across different channel conditions. As shown in Fig.~\ref{fig:semantic_gap_all}, when $\gamma_{\mathrm{eve}} = 0$, corresponding to the non-adversarial baseline, the semantic accuracy at the destination and the relay are nearly identical. This indicates negligible protection against eavesdropping and confirms that, without adversarial training, the learned representation is equally informative to both receivers (namely, the destination and the relay).

As $\gamma_{\mathrm{eve}}$ increases, a clear destination-relay separation emerges across all SNRs. The relay's semantic accuracy decreases significantly, while the destination remains stable initially. This shows that adversarial loss reduces semantic separability at the relay, degrading the relay's inference capability while preserving representation utility for the destination.

At moderate values of $\gamma_{\mathrm{eve}}$, the destination-relay accuracy gap reaches its maximum, corresponding to an effective trade-off between utility and privacy. In this regime, strong protection is achieved with only limited degradation in the destination's performance. The consistency of this behavior across SNR conditions suggests that adversarial training is robust and effective in controlling semantic leakage.

As $\gamma_{\mathrm{eve}}$ increases further, the destination's accuracy begins to degrade more noticeably, while the accuracy gap saturates or slightly decreases. This indicates that excessively strong adversarial pressure distorts the latent representation, affecting both receivers and reducing overall semantic discriminability rather than selectively suppressing the relay.

SNR also affects these trends. At higher SNR, the destination remains stable across a wider range of $\gamma_{\mathrm{eve}}$, yielding a larger accuracy gap. At lower SNR, the system becomes more sensitive to adversarial representation shaping, and the destination's accuracy degrades more rapidly as $\gamma_{\mathrm{eve}}$ increases. Nevertheless, meaningful destination-relay separation is still achieved.

Overall, the results demonstrate that the proposed iterative adversarial training framework effectively enlarges the destination-relay semantic accuracy gap both over training time and as a function of the adversarial loss weight. The framework enables controlled navigation of the privacy-utility trade-off, achieving strong semantic protection while preserving performance along the legitimate communication path.

\subsection{Reconstruction Quality}

In addition to semantic accuracy, we evaluate the impact of the adversarial training on reconstruction quality at both destination and relay. While the iterative framework is designed to enlarge the semantic accuracy gap, an important question is whether this degradation at the relay is accompanied by noticeable distortion in the reconstructed sample. To quantify this, we measure the difference between the destination and the relay in terms of PSNR and SSIM across SNR conditions.

\begin{table}[t]
\centering
\caption{Destination-relay reconstruction quality gap.}
\footnotesize
\label{tab:psnr_ssim_gap}
\begin{tabular}{c|cc|cc}
\hline
\multirow{2}{*}{SNR (dB)} & \multicolumn{2}{c|}{PSNR Difference (dB)} & \multicolumn{2}{c}{SSIM Difference} \\ 
 & Max & Avg & Max & Avg \\
\hline
-6 & 0.38 & 0.24 & 0.018 & 0.008 \\
0  & 0.54 & 0.28 & 0.014 & 0.007 \\
6  & 0.34 & 0.23 & 0.010 & 0.008 \\
\hline
\end{tabular}
\end{table}

The results in Table~\ref{tab:psnr_ssim_gap} show that the difference in reconstruction quality between the destination and the relay remains consistently small across all SNR values. Both the maximum and average PSNR differences are limited, indicating that the relay is still able to reconstruct the input sample with fidelity comparable to that of the destination. Similarly, the SSIM differences remain very small, confirming that perceptual structural similarity is largely preserved at the relay.

These results demonstrate stealthy protection, where the relay's semantic accuracy is degraded while reconstruction quality remains largely intact. Thus, the relay observes structurally consistent signals but fails to infer their semantic content, achieving semantic-specific degradation without noticeable signal distortion. Overall, a large semantic accuracy gap with minimal reconstruction difference indicates effective selective information hiding.

\section{Conclusion \label{sec:conclusion}}
This paper addressed the privacy implications of relay-assisted SemCom and showed that relay nodes can act as capable eavesdroppers, extracting semantic meaning from transmitted latent representations. We first demonstrated that the relay can achieve semantic inference and reconstruction performance comparable to that of the legitimate receiver, revealing a fundamental vulnerability. To mitigate this issue, we proposed an iterative adversarial training framework that explicitly accounts for a strong adaptive eavesdropper. By alternating between optimizing the legitimate system and the eavesdropper, the framework learns representations that preserve utility at the intended receiver while degrading semantic inference at the relay. This significantly enlarges the semantic accuracy gap across channel conditions while maintaining high reconstruction fidelity. The protection is achieved in a stealthy manner, selectively suppressing semantic information without noticeable signal distortion. These findings highlight the privacy risks of semantic relays and the need for privacy-aware training in relay-assisted SemCom systems.

Future research directions include (i) extending relay-assisted SemCom to multi-relay, multi-hop, and multi-user networks to study semantic leakage in distributed cooperative architectures, (ii) developing cross-layer privacy-aware SemCom designs that jointly optimize routing, scheduling, relay selection, and semantic resource allocation while accounting for leakage across communication paths, and (iii) extending the framework to decentralized and federated semantic learning settings where privacy leakage may arise through shared model updates and latent representations.

\bibliographystyle{IEEEtran}
\bibliography{references}

\end{document}